\documentclass[num-refs]{wiley-article}

\usepackage{siunitx}
\usepackage{float}
\usepackage{ifpdf}
\papertype{Original Article}
\paperfield{Journal Section}

\title{\textcolor{black}{Observation of Shear Strain in Ion-Implanted Diamond Substrate and Diamond Nanophotonic Structures}}

\abbrevs{}

\author[1\authfn{1}]{Ayan Majumder}
\author[3\authfn{1}]{Vivek K Shukla}
\author[2,1]{Anuj Bathla}
\author[4]{Brajesh S. Yadav}
\author[4]{Nanhey Singh}
\author[3\authfn{2}]{Padmnabh Rai}
\author[1\authfn{3}]{Kasturi Saha}

\contrib[\authfn{1}]{Equally contributing authors.}

\affil[1]{Department of Electrical Engineering, Indian Institute of Technology Bombay, Powai, Mumbai-400076, India}
\affil[2]{Center for Research in Nano Technology and Science, Indian Institute of Technology Bombay, Powai, Mumbai-400076, India}
\affil[3]{School of Physical Sciences, UM-DAE Centre for Excellence in Basic Sciences, University of Mumbai, Mumbai-400098, India}
\affil[4]{Solid State Physics Laboratory, Lucknow Road, Timarpur, Delhi-110054, India}
\corraddress{Department of Electrical Engineering, Indian Institute of Technology Bombay, Powai, Mumbai-400076, India}
\corremail{kasturis@ee.iitb.ac.in, 194076009@iitb.ac.in}

\presentadd[\authfn{2}]{}

\fundinginfo{}

\runningauthor{}

\begin{document}

\begin{frontmatter}
\maketitle

\begin{abstract}
Negatively charged nitrogen-vacancy (NV) centers and other color centers in diamonds have emerged as promising platforms for quantum communication, quantum information processing, and nanoscale sensing, owing to their long spin coherence times, fast spin control, and efficient photon coupling. Deterministic placement of individual color centers into nanophotonic structures is critical for scalable device integration, and ion implantation is the most viable technique. Nanofabrication processes, including diamond etching, are essential to realize these structures but can introduce crystal strain through lattice damage. In this work, we investigate the impact of ion implantation and nanofabrication-induced strain on the electronic spin levels of NV-centers. We demonstrate that the zero-field continuous-wave optically detected magnetic resonance (CW-ODMR) spectroscopy serves as a sensitive probe of local crystal strain. \textcolor{black}{We report the presence of a shear strain feature in diamond substrates arising from the ion-implantation and nanofabrication processes, as evidenced by the asymmetric splitting} \textcolor{black}{observed in the zero-field CW-ODMR spectrum of NV-centers.} 

% Please include a maximum of seven keywords
\keywords{\hbox{Nitrogen-vacancy center}, \hbox{nanophotonics}, \hbox{crystal-strain}, \hbox{ion-implantation}}
\end{abstract}
\end{frontmatter}

%%%%%%%%%%%%%%%%%%%%%%%%%%%%%%%%%%%%%%%%%%%%%%%%%%%%%%%%%%%%%%%%%%%%%%%%
%\linenumbers 
\section{Introduction}
\label{section-1}
 Color defect centers in diamond are promising candidates for quantum communication \cite{ruf2021quantum}, quantum information processing \cite{suter2017single,wolfowicz2021quantum}, and nanoscale sensing \cite{du2024single,bucher2019quantum} applications due to their long coherence times, fast spin-state manipulation, and the ability to be optically initialized and read out \cite{pezzagna2021quantum}. Among these, the NV-center is particularly notable, as it can be optically initialized and read out even at room temperature \cite{pezzagna2021quantum}. The electronic spin states of NV-centers are highly sensitive to variations in electric and magnetic fields, temperature, and local strain in the host diamond lattice \cite{levine2019principles}. Owing to this sensitivity, single NV-centers can be incorporated into diamond nanopillars or nanodiamonds, serving as atomic-scale probes for super-resolution scanning probe microscopy \cite{rondin2014magnetometry} or targeted in vivo magnetic noise sensing \cite{modak2023few} in a cell environment \cite{mzyk2022relaxometry}. Additionally, thin layers (a few nanometers thick) of high-density NV-centers can be used for magnetic imaging of samples, offering a wide field of view, micrometer-scale resolution \cite{levine2019principles,parashar2022sub}, and high magnetic field sensitivity \cite{barry2020sensitivity}.\\
For all such applications, improving light-matter interaction \cite{janitz2020cavity,majumder2022engineering,knaut2024entanglement,reiserer2015cavity} and efficient fluorescence collection \cite{baier2020orbital} is essential to enhance the fidelity of quantum operations \cite{ruf2021quantum,pfaff2013demonstration}. In particular, improving the fidelity of quantum state transfer from photonic qubits to spin qubits requires enhanced light-matter interaction, which can be achieved by placing the NV-center in a high-cooperativity optical cavity \cite{knaut2024entanglement}. Ion implantation is a promising technique for positioning single emitters at desired locations within nanophotonic cavities \cite{ruf2021quantum,knaut2024entanglement,harris2025high}. Similarly, advanced nanofabrication techniques are crucial for the development of photonic structures that enhance both light-matter interaction and photon collection efficiency \cite{marseglia2018bright,mccloskey2020enhanced}.\\
However, both ion implantation and nanofabrication introduce lattice damage that results in crystal strain within the diamond substrate \cite{liang2025strain,alam2024determining,udvarhelyi2018spin}. Several methods exist for strain characterization, including those based on negatively-charged silicon vacancy (SiV) centers \cite{bates2021using,klotz2025ultra}, NV-center zero-phonon line shifts \cite{liang2025strain,grazioso2013measurement}. A widely used technique is zero-field ODMR \cite{grazioso2013measurement}. Previous studies have used ODMR to measure axial and non-axial strain components using preferentially aligned NV-centers in unprocessed polycrystalline diamond and nanocrystals containing few NV-centers, or by analyzing the splitting of single-NV ODMR signals \cite{trusheim2016wide,shukla2025anti}. \textcolor{black}{ODMR spectroscopy has also been utilized to probe strain components in diamond nanophotonic structures and waveguides, as demonstrated by Sebastian Knauer et al. \cite{knauer2020situ}, and M. Sahnawaz Alam et al. \cite{alam2024determining}, respectively. In the work by Sebastian Knauer et al. \cite{knauer2020situ}, diamond nanophotonic structures were fabricated using a focused ion beam milling process, and ODMR spectroscopy was used to investigate the resulting local strain within these structures. In contrast, M. Sahnawaz Alam et al. \cite{alam2024determining} fabricated diamond waveguides through femtosecond laser writing, where tightly focused laser pulses induced structural modifications—such as amorphization and graphitization—leading to strain in the crystal lattice. Both studies reported asymmetric splitting in the ODMR spectrum of NV centers. Furthermore, M. Sahnawaz Alam et al. \cite{alam2024determining} developed a theoretical model to explain the origin of the asymmetric zero-field splitting observed in the ODMR spectrum.}\\
\textcolor{black}{In this study, we prepared two different samples. In the first sample, nitrogen ion implantation was performed on a chemical vapor deposition (CVD)-grown diamond, followed by annealing to create NV-centers. In the second sample, diamond nanopillar structures were fabricated on a CVD-grown diamond with a low concentration of NV-centers. Zero-field CW-ODMR spectroscopy was employed to probe the local strain induced by the ion-implantation and nanofabrication processes in both samples, using NV-centers as atomic-scale sensors. For both samples, we observed an asymmetric splitting of the zero-field ODMR spectrum, which arises from the in-plane shear strain developed in the diamond lattice.}\\
This paper is organized as follows. In Subsection~\ref{subsection-2.0}, we describe the SRIM (Stopping and Range of Ions in Matter) simulations and the sample preparation process for ion-implanted single-crystal CVD-grown diamond, including Three-Dimensional Finite-Difference Time-Domain (3D-FDTD) simulations implemented to optimize the nanopillar dimensions and details of the fabrication process. Subsection~\ref{subsection-2.1} presents Raman spectroscopy measurements for characterizing strain in the diamond crystal. In Subsection~\ref{subsection-2.2}, we describe the measurement of magnetic resonance spectra for NV-centers in both ion-implanted diamond and diamond nanophotonic structures. \textcolor{black}{Subsection~\ref{subsection-2.3} discusses the origin of the asymmetric splitting caused by shear strain induced by the ion-implantation and nanofabrication processes in the diamond samples. The theoretical model proposed by M. Sahnawaz Alam et al. \cite{alam2024determining} is briefly reviewed to explain the observed strain characteristics.} We conclude in Section~\ref{section-3}, and Section~\ref{section-4} provides details of our experimental setups for Raman spectroscopy and the microwave-integrated laser scanning confocal microscopy system.

%%%%%%%%%%%%%%%%%%%%%%%%%%%%%%%%%%%%%%%%%%%%%%%%%%%%%%%%%%%%%%%%%%%%%%%%

\section{Results and Discussion}
\label{section-2}

\textcolor{black}{In this section, the sample preparation details, along with the measurement procedures for Raman spectroscopy and zero-field ODMR spectroscopy, are presented, and the observed results are discussed. Two different diamond samples were investigated. The first sample is a nitrogen-ion-implanted CVD-grown diamond. The second sample consists of diamond nanopillar structures fabricated on a CVD-grown diamond with a low concentration of NV-centers. Zero-field CW-ODMR spectroscopy was implemented to probe the local strain induced by the ion-implantation and nanofabrication processes in both samples, with NV-centers serving as atomic-scale sensors.}

\subsection{Sample Preparation}
\label{subsection-2.0}
\subsubsection{SRIM Simulation and Nitrogen-ion Implantation}
%\label{subsubsection-2.0.1}

\begin{figure}[H]
\centering
\includegraphics[width=12cm]{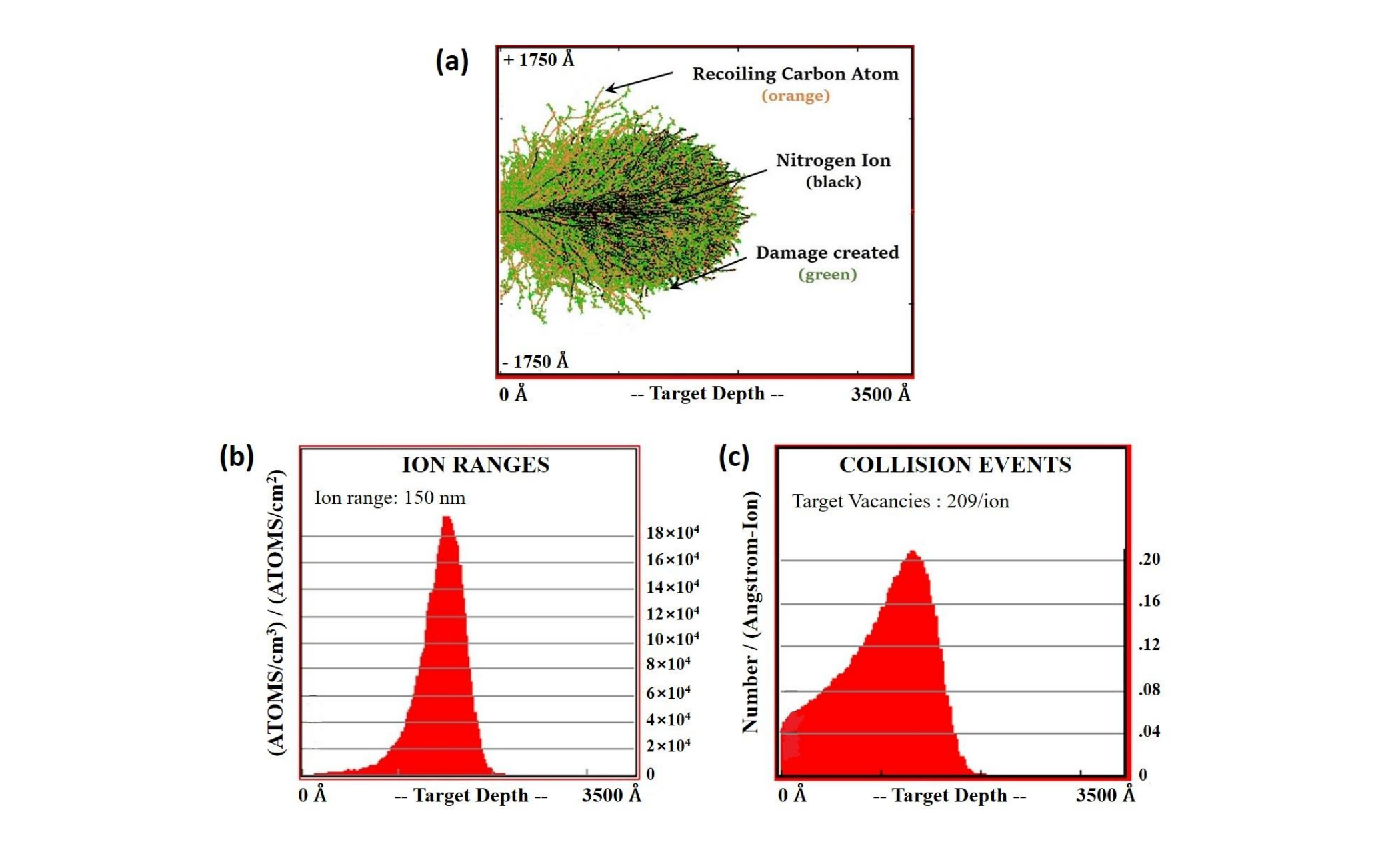}
\vspace*{-10pt}
\caption{Results of the SRIM simulation of nitrogen-implanted diamond conducted at energy 130 keV. (a) Ion trajectories of nitrogen ions, recoiled carbon atoms, and created vacancies are depicted with different colors, (b) ion distribution profile, and (c) target carbon vacancies, as a function of depth.}
\label{fig1}
\end{figure}

The single-crystal diamond (SCD) samples were grown on $(100)$-oriented (Type IIa) diamond substrates ($7.5$ mm $\times 7.2$ mm $\times 0.8$ mm, in-house CVD-grown SCD with nitrogen concentration $< 100$ ppb) using a custom-built microwave plasma CVD reactor ($2.45$ GHz, $6$ kW). The optimal growth conditions for homoepitaxial SCDs are as follows: Microwave power $5000$ W, CH$_4/$H$_2$ flow ratio $10\%$, N$_2/$H$_2$ flow ratio $0.004\%$, substrate temperature $~950^{\circ}$C, and chamber pressure $~125$ Torr. After growth, the samples underwent laser cutting, mechanical polishing, and cleaning using a boiling ternary acid mixture (HNO$_3 :$ HClO$_4 :$ H$_2$SO$_4 = 1:1:1$ by volume) to selectively etch away graphitic carbon from the diamond surface. The CVD-grown diamond samples (labeled DRM$-8$, $9$, and $10$) were then employed for nitrogen ion implantation. The sample dimensions were $5.0$ mm $\times 5.4$ mm $\times 0.4$ mm, $4.8$ mm $\times 5.1$ mm $\times 0.4$ mm, and $6.0$ mm $\times 5.0$ mm $\times 0.3$ mm for DRM$-8$, DRM$-9$, and DRM$-10$, respectively. 
Ion implantation was carried out using nitrogen ions ($^{14}N^+$, energy $~130$ keV) at room temperature with three different doses: $1\times10^{13}$ (DRM$-8$), $1\times10^{14}$ (DRM$-9$), and $2\times10^{16}$ (DRM$-10$) ions$/$cm$^2$, respectively, utilizing IBS, IMC$200$ ion implanter. A homogeneous circular ion beam of diameter $~1$ mm was rastered over the sample to achieve uniform irradiation at the desired dose. The samples were subsequently annealed at $~1000^{\circ}$C in an argon (Ar) atmosphere for $2$ h at a chamber pressure of $~5\times10^{-3}$ mbar to restore the implantation-induced lattice damages. Additionally, the samples were thoroughly acid-cleaned to eliminate any carbon or graphitic deposits from the surface before conducting any measurements.\\
Monte-Carlo simulations were performed using the stopping and range of ions in matter (SRIM) software (version 2008.04) to estimate the concentration of implantation-induced vacancies in the diamond matrix \cite{ziegler2008srim}. The estimated trajectory of nitrogen ions and recoiled carbon atoms resulting from collisions in the diamond matrix is depicted in Figure \ref{fig1}(a). A nearly Gaussian distribution of nitrogen ions in the diamond matrix shows a peak concentration at a depth of $~150$ nm below the top surface (Figure \ref{fig1}(b)). The vacancy density, signifying the extent of lattice damage caused by the displacement of carbon atoms from their lattice sites, is shown in Figure \ref{fig1}(c). Table $1$ summarizes the implantation energy, dose, and corresponding estimated damage (vacancies/cm$^3$) for the nitrogen-implanted diamond samples. 
\begin{table}[H]
\caption{SRIM simulation results for vacancy concentration at different ion fluences in the diamond matrix.}
\begin{threeparttable}
\begin{tabular}{lccrr}
\headrow
\thead{Sample name} & \thead{Implantation Energy (keV)} & \thead{Ion Fluence/Dose (ions/cm$^2$)} & \thead{Damage
(vacancies/cm$^3$)}\\
DRM-8 & 130 & $1\times 10^{13}$ & $2.08\times 10^{20}$\\
DRM-9 & 130 & $1\times 10^{14}$ & $2.09\times 10^{21}$\\
DRM-10 & 130 & $1\times 10^{16}$ & $2.15\times 10^{23}$\\
\hline  % Please only put a hline at the end of the table
\end{tabular}
\begin{tablenotes}
\item 
\end{tablenotes}
\end{threeparttable}
\end{table}

%%%%%%%%%%%%%%%%%%%%%%%%%%%%%%%%%%%%%%%%%%%%%%%%%%%%%%%%%%%%%%%%%%%%%%%%

\subsubsection{3D-FDTD-Simulation and Fabrication of Diamond Nanopillars}
\label{subsubsection-2.0.1}

\begin{figure}[H]
\centering
\includegraphics[width=12cm]{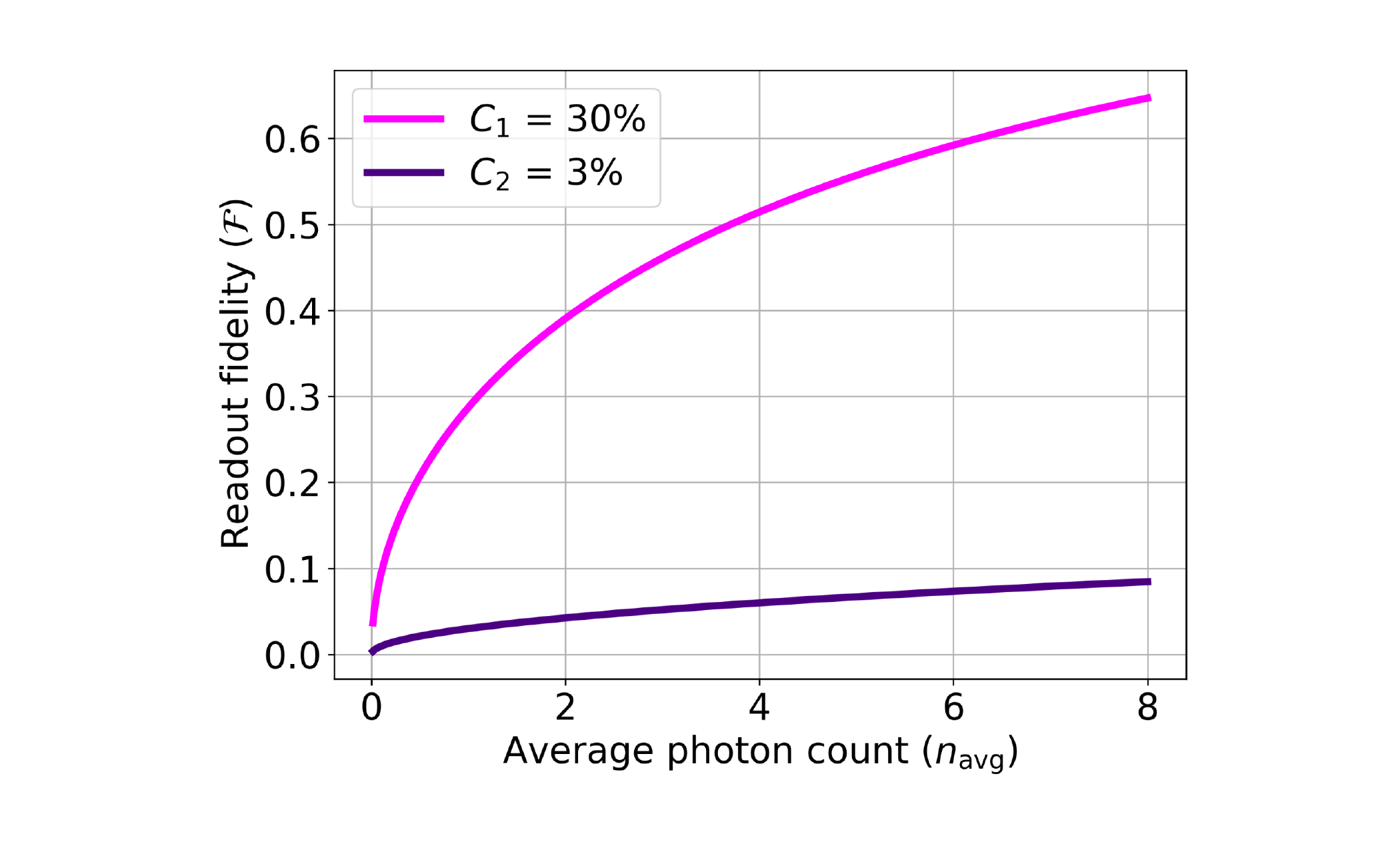}
\vspace*{-20pt}
\caption{ \textcolor{black}{Readout fidelity as a function of $n_{\mathrm{avg}}$ for both a single NV-center and an ensemble of NV-centers.} $C_1$ and $C_2$ represent the fluorescence contrast for the single NV-center and the ensemble, respectively. The readout duration for this plot is 300~ns.}
\label{fig2-0}
\end{figure}

\begin{figure}[H]
\centering
\includegraphics[width=12cm]{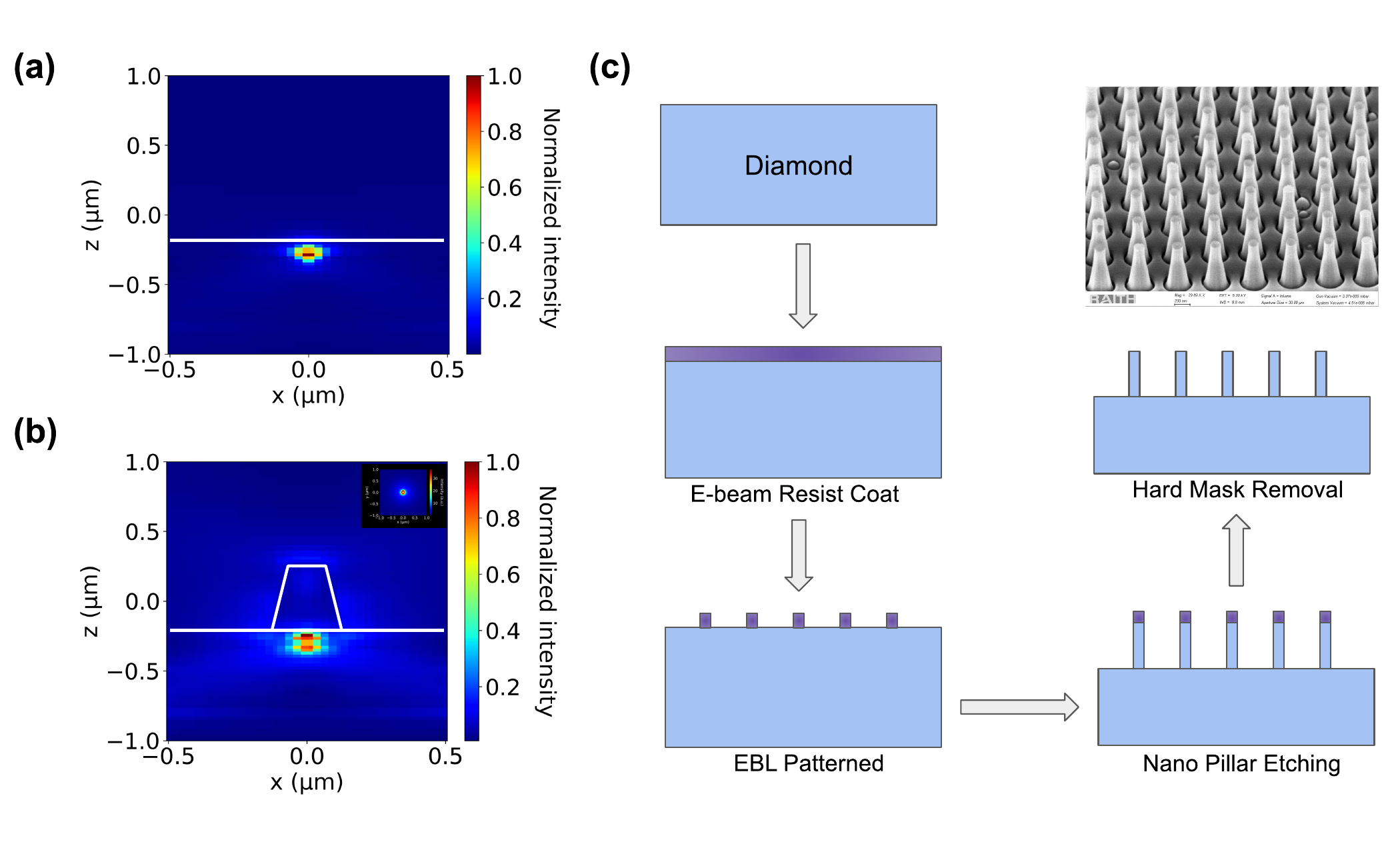}
%\vspace*{-15pt}
\caption{(a) Simulated normalized intensity distribution of a dipole emitter inside a diamond slab. (b) Simulated normalized intensity distribution of a dipole emitter with a diamond nanopillar. The inset figure shows the intensity distribution in the XY plane, which is around $600$ nm above the dipole emitter. (c) Schematic of the fabrication process flow of the diamond nanopillar sample. The SEM image of the fabricated diamond nanopillar sample is shown here.}
\label{fig2}
\end{figure}

Improving the photon collection efficiency is essential for quantum sensing \cite{mccloskey2020enhanced} and quantum network \cite{ruf2021quantum} applications. The readout fidelity determines the sensitivity as well as the spin-spin entanglement generation rate based on single-photon detection. The $n_{avg}$ and the contrast ($C$) determine the readout fidelity ($\mathcal{F}$) \cite{barry2020sensitivity},
\begin{equation}    
\mathcal{F} = \frac{C\sqrt{n_{avg}}}{\sqrt{C^2 n_{avg}+1}}
\end{equation}
The average number of photons collected per NV-center per measurement is denoted by $n_{\mathrm{avg}}$. Improving the readout fidelity necessitates increasing $n_{\mathrm{avg}}$, which is directly enhanced by the photon collection efficiency. \textcolor{black}{Figure~\ref{fig2-0} presents the readout fidelity as a function of $n_{\mathrm{avg}}$ for both a single NV-center and an ensemble of NV-centers. To obtain a reasonable estimate of the average photon counts, a readout duration of $300$~ns is considered for this analysis.} The high refractive index of diamond (approximately 2.41) causes total internal reflection confinement, preventing photons from being efficiently collected. The calculated collection efficiencies for photons emitted directly through the $\{100\}$ diamond surface are 3.7$\%$ and 10.4$\%$ for air and oil-immersion objectives, respectively, with numerical apertures of 0.95 and 1.49 \cite{barry2020sensitivity}.  Modifying the diamond surface is helpful in increasing photon collection efficiency.
\textcolor{black}{For the fabrication of diamond nanopillars, a CVD-grown single-crystal diamond with a (100) surface orientation and a low concentration of NV-centers was used. To evaluate the performance of the nanophotonic structures, three-dimensional finite-difference time-domain (3D-FDTD) simulations were carried out to determine the optimal dimensions of the diamond nanopillars for enhanced light collection efficiency \cite{marseglia2018bright,mccloskey2020enhanced}.}\\
\textcolor{black}{Considering a single NV-center whose symmetry axis is oriented with respect to the laboratory frame by the polar and azimuthal angles $\theta$ and $\phi$, respectively \cite{reuschel2022vector,zhu2024visualized}, the unit vector along the NV-axis can be expressed as}
\begin{equation}
   \hat{e}_{NV} = \left( \begin{array}{cc} \sin\theta\cos\phi \\
           \sin\theta\sin\phi \\
           \cos\theta \end{array} \right) 
\end{equation}
The two mutually orthogonal transition dipoles associated with the NV center, denoted by $\vec{d}_1$ and $\vec{d}_2$ with dipole moment $d$, are given by \cite{reuschel2022vector,zhu2024visualized}, 
\begin{equation}
   \vec{d}_{1} = d \left( \begin{array}{cc} -\sin\phi \\
           \cos\phi \\
           0 \end{array} \right), 
%\end{equation}
%\begin{equation}
   \vec{d}_{2} = d \left( \begin{array}{cc} \cos\theta\cos\phi \\
           \cos\theta\sin\phi \\
           -\sin\theta \end{array} \right) 
\end{equation}
For our case, the surface orientation of the diamond sample is (100), and the corresponding four possible NV-center orientations and the associated dipole orientations are listed in Table~2.
\begin{table}[]
\caption{NV-axis and dipole vector orientations with corresponding spherical angles.}
\begin{center}
\begin{threeparttable}
\begin{tabular}{lccc}
\headrow
\thead{NV Orientation} & \thead{Vector} & \thead{Cartesian Components} & \thead{Spherical Angles ($^\circ$)} \\
NV-1 & $\vec{e}_{\text{NV}}$ & $[-0.229,\ -0.805,\ -0.547]$ & $\theta = 123.14,\ \phi = 254.14$ \\
     & $\vec{d}_1$           & $[-0.962,\ 0.273,\ 0.0]$     & $\theta = 90.00,\ \phi = 164.14$ \\
     & $\vec{d}_2$           & $[0.149,\ 0.526,\ -0.837]$   & $\theta = 146.86,\ \phi = 74.14$ \\
NV-2 & $\vec{e}_{\text{NV}}$ & $[-0.773,\ 0.239,\ 0.588]$   & $\theta = 53.98,\ \phi = 162.81$ \\
     & $\vec{d}_1$           & $[0.296,\ 0.955,\ -0.0]$     & $\theta = 90.00,\ \phi = 72.81$ \\
     & $\vec{d}_2$           & $[-0.562,\ 0.174,\ -0.809]$  & $\theta = 143.98,\ \phi = 162.81$ \\
NV-3 & $\vec{e}_{\text{NV}}$ & $[0.238,\ 0.766,\ -0.597]$   & $\theta = 126.64,\ \phi = 72.75$ \\
     & $\vec{d}_1$           & $[0.955,\ -0.297,\ 0.0]$     & $\theta = 90.00,\ \phi = 342.75$ \\
     & $\vec{d}_2$           & $[-0.177,\ -0.57,\ -0.802]$  & $\theta = 143.36,\ \phi = 252.75$ \\
NV-4 & $\vec{e}_{\text{NV}}$ & $[0.813,\ -0.236,\ 0.533]$   & $\theta = 57.80,\ \phi = 343.79$ \\
     & $\vec{d}_1$           & $[-0.279,\ -0.96,\ 0.0]$     & $\theta = 90.00,\ \phi = 253.79$ \\
     & $\vec{d}_2$           & $[0.512,\ -0.149,\ -0.846]$  & $\theta = 147.8,\ \phi = 343.79$ \\
\hline
\end{tabular}
\begin{tablenotes}
\item 
\end{tablenotes}
\end{threeparttable}
\end{center}
\end{table}
\textcolor{black}{Based on the diamond sample, two simulation scenarios were considered. In the first case, all possible dipole orientations corresponding to the four crystallographic orientations of the NV-center \cite{reuschel2022vector} were placed inside the diamond crystal with a center emission wavelength of $637$ nm. The dipole emitters were positioned approximately $10$ nm below the diamond surface.}\\
\textcolor{black}{In the second case, a truncated conical-shaped diamond nanopillar was introduced on the crystal surface. The refractive index of diamond was taken to be $2.418$ for both configurations \cite{majumder2022engineering}. An air medium with a refractive index of $1.0$ was assumed to surround the nanostructure. The nanopillar height was $576$ nm, while the diameters of the top and bottom surfaces were approximately $146$ nm and $304$ nm, respectively. A perfectly matched layer (PML) boundary condition was applied around the simulation domain to suppress reflections from the simulation boundaries.}\\
\textcolor{black}{Figures~\ref{fig2}(a) and (b) present the simulated normalized intensity distributions at a wavelength of $637$ nm, with the inset showing the intensity profiles collected by the monitoring plane. The simulation results indicate enhanced confinement and directional emission of photoluminescence (PL) when the dipole emitter is located near the diamond nanopillar. Compared with the pristine diamond crystal containing an ensemble of emitters close to the surface, the emission from the nanopillar structure becomes more directional and better focused. Further simulation details are mentioned in the supporting information section (see Figure~\ref{figS1}).}\\
The diamond pillar fabrication was carried out based on the simulated parameters. Figure~\ref{fig2}(c) presents a flowchart of the fabrication process. Initially, the diamond substrate was cleaned using a tri-acid solution. A bilayer resist (495KA4 and 950KA2 PMMA) was then spin-coated onto the substrate and baked at $180^\circ$C. Electron beam lithography (Raith-150Two) was used for patterning. Subsequently, a $60$ nm thick titanium (Ti) layer was deposited using a sputtering technique. The diamond etching was performed using inductively coupled plasma reactive ion etching (ICP-RIE), following the defined pattern. Finally, hydrofluoric acid (HF) was used to remove the residual Ti-mask. \textcolor{black}{The diamond crystal was etched (in ICP-RIE system) for 3~min with 99~s.c.c.m. of oxygen gas, 1000~W bias power, 85~W RF-power (-200~V DC bias) at a chamber pressure of 1~pascal}. The scanning electron microscopy (SEM) image of the fabricated diamond nanopillar array is shown in Figure~\ref{fig2}(c).

%%%%%%%%%%%%%%%%%%%%%%%%%%%%%%%%%%%%%%%%%%%%%%%%%%%%%%%%%%%%%%%%%%%%%%%%

\subsection{Raman Spectroscopy}
\label{subsection-2.1}

\begin{figure}[H]
\centering
\includegraphics[width=12cm]{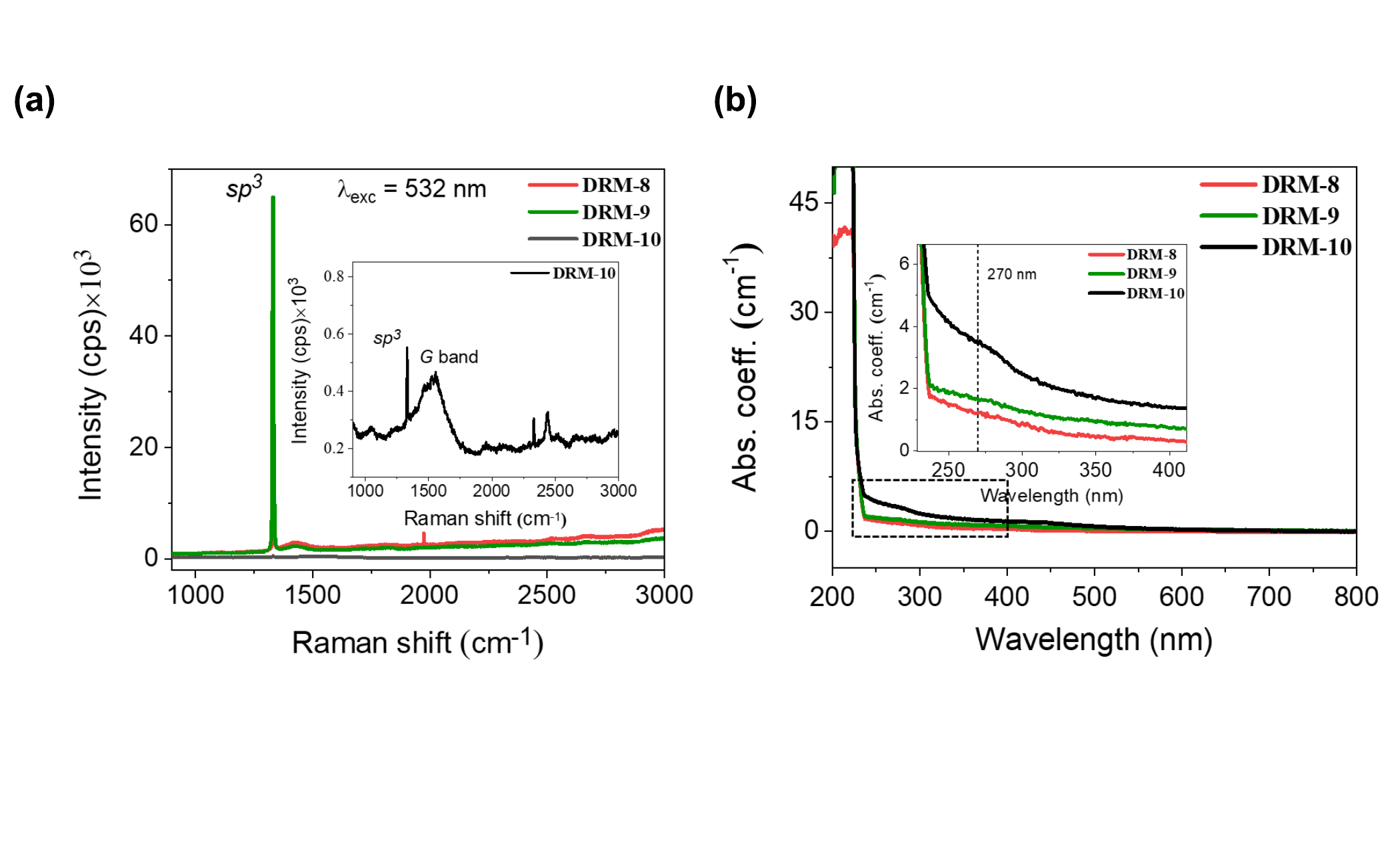}
\vspace*{-43pt}
\caption{(a) Raman and corresponding (b) post-annealing UV-visible absorption spectra of the diamond samples. The spectra shown in the insets are their enlarged views.}
\label{fig3}
\end{figure}

Raman measurements were performed to determine the distribution and evolution of the diamond Raman peak position and the defect peaks that arise from sample irradiation. The Raman signature of the implanted samples is evident by the narrow one-phonon peak that appeared at $~1332$ cm$^{-1}$, with no significant background observed. Upon implantation, the one-phonon peak broadens and shifts to higher wave numbers, indicating damage to the diamond lattice. The damage caused by ion implantation, at $1\times10^{13}$ and $1\times10^{14}$ ions/cm$^2$ fluences, remains below the critical graphitization threshold ($~2\times10^{22}$ vacancies/cm$^3$), which allows the sp$^3$-bonding to be preserved in the samples DRM-8 and DRM-9 after annealing \cite{uzan1995damage,jimenez2023boron}. High fluences ($1\times10^{16}$ ions/cm$^2$) cause significant lattice damage in the diamond sample (DRM-10), as evidenced by the weakened diamond Raman peak intensity (Fig. \ref{fig3}(a), inset). Since the intensity of the diamond peak correlates with the volume of regular diamond within the laser spot, its reduction can serve as an estimate of the degree of amorphization \cite{jimenez2023boron}. Annealing leads to a slight reduction in the damaged region but does not significantly enhance the intensity of the diamond Raman peak, instead resulting in a graphite-related G peak at $~1580$ cm$^{-1}$. Figure \ref{fig3}(b) presents the room temperature UV-visible absorption spectra ($200 - 800$ nm) of ion-implanted diamond samples. The DRM-10 sample shows notably stronger absorption at $~270$ nm (Fig. \ref{fig3}(b), inset), which correlates with its higher substitutional nitrogen ($N_s^0$) concentration. The $N_s^0$ concentration was determined by fitting the UV absorption spectrum (300-450 nm region) with a second-order polynomial \cite{shukla2024influence,nistor2000nitrogen}, using [$N_s^0$] (ppm) $= 0.56 \times \alpha_{270}$, where $\alpha_{270}$ (cm$^{-1}$) is the absorption coefficient at $270$ nm. The estimated nitrogen concentrations were $300$ ppb (DRM-8), $250$ ppb (DRM-9), and $760$ ppb (DRM-10).

%%%%%%%%%%%%%%%%%%%%%%%%%%%%%%%%%%%%%%%%%%%%%%%%%%%%%%%%%%%%%%%%%%%%%%%%

\subsection{Magnetic Resonance Spectroscopy of NV-Centers in Bulk Diamond Crystal and Nanophotonic Structure}
\label{subsection-2.2}

\begin{figure}[H]
\centering
\includegraphics[width=12cm]{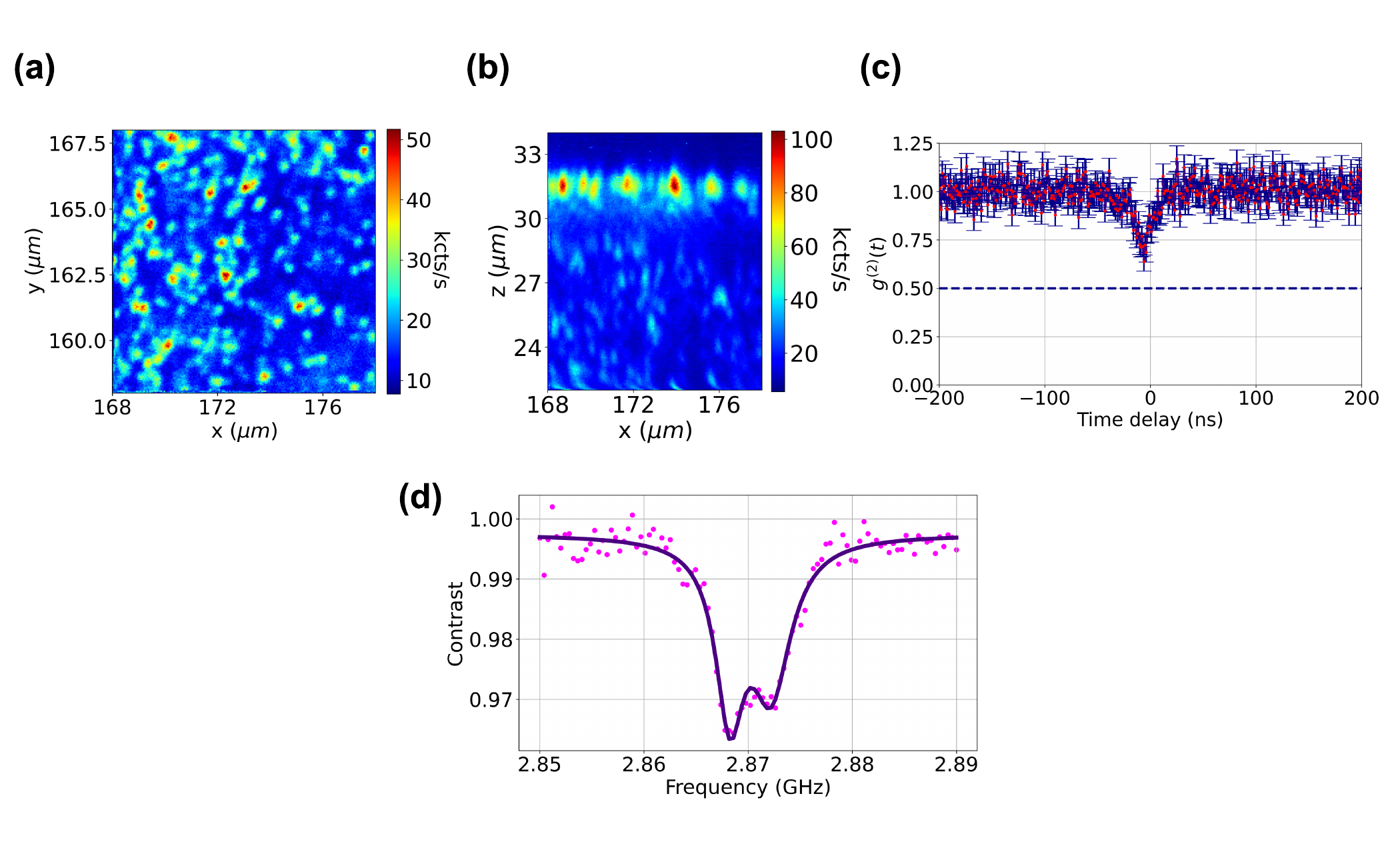}
\vspace*{-10pt}
\caption{(a) XY-confocal scan of the nitrogen-ion-implanted sample  \textcolor{black}{(DRM-8)}. (b) XZ-confocal scanning of the nitrogen-ion-implanted sample \textcolor{black}{(DRM-8)}. (c) Second-order correlation measurement data at a diffraction-limited spot. (d) Zero-field ODMR spectrum at the same diffraction-limited spot. Two Lorentzian profiles were used to fit the experimental data. There was an imbalance between the two Lorentzian peaks in the zero-field CW-ODMR data.}
\label{fig4}
\end{figure}

\begin{figure}[H]
\centering
\includegraphics[width=12cm]{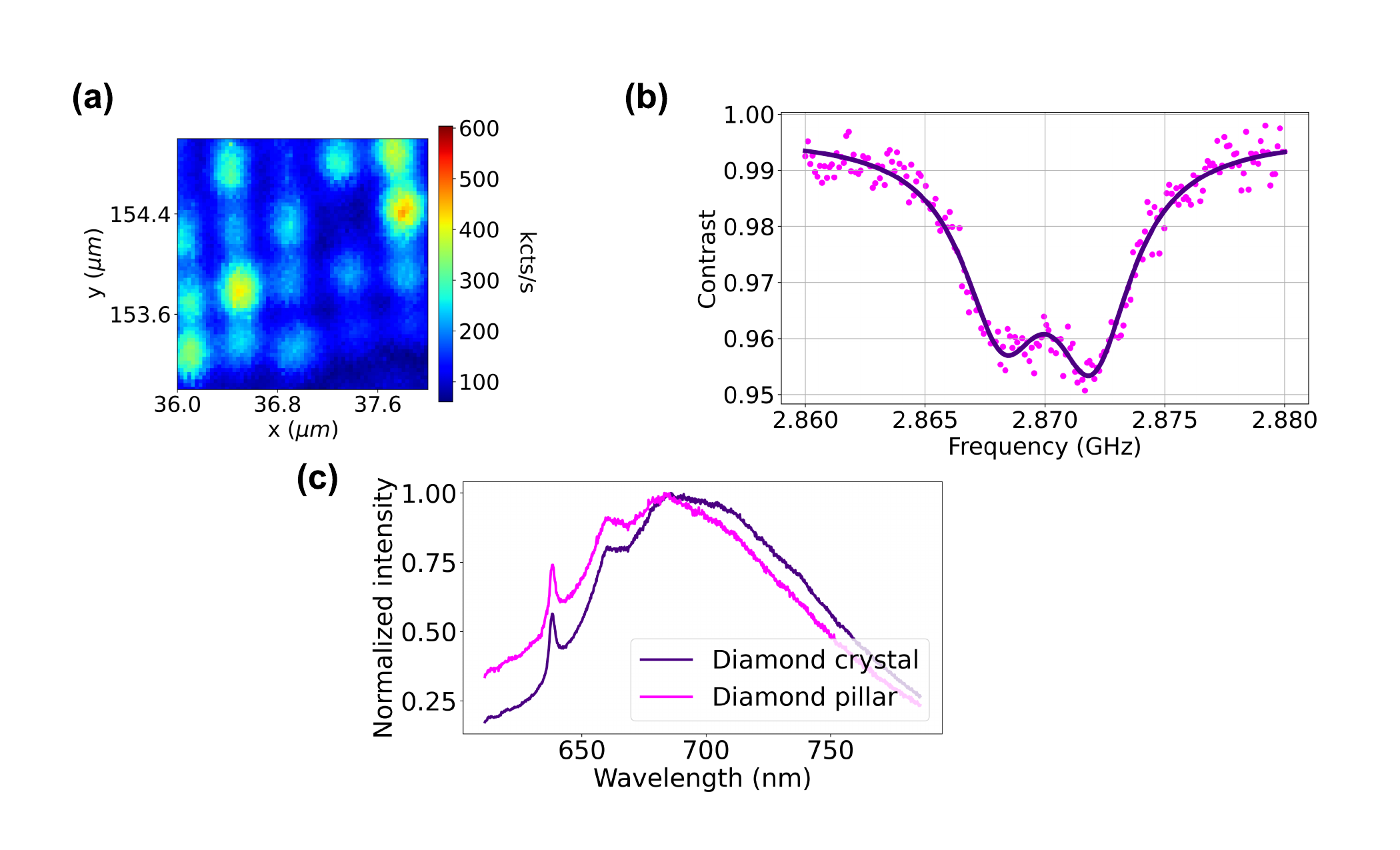}
\vspace*{-10pt}
\caption{(a) Confocal image of the diamond nanopillar array. (b) Zero-field ODMR spectrum of diamond nanopillar. Two Lorentzian profiles are considered to fit the experimental data. There is an imbalance between the two Lorentzian peaks in the zero-field CW-ODMR data. (c) \textcolor{black}{Photoluminescence spectroscopy data for the pristine and diamond nanopillar samples in the presence of 532~nm laser excitation}.}
\label{fig5}
\end{figure}

A home-built laser-scanning confocal microscope with a microwave excitation module was used for confocal imaging of the bulk diamond sample and spectroscopy of the color centers in the diamond. Figures~\ref{fig4}(a), (b), and~\ref{fig5}(a) show confocal images of the samples. The bright spots are clusters of NV-centers in the diffraction-limited spot. In Figure~\ref{fig5}(a), the photon counts from the diamond nanopillars are five times higher than the background counts. The PL enhancement was consistent with the simulated scenario. Figure~\ref{fig5}(c) shows the \textcolor{black}{PL-spectroscopy} data. Figure~\ref{fig4}(c) shows the second-order correlation measurement, which confirms the number of emitters at the diffraction-limited spot. For both samples, the number of NV-centers at the diffraction-limited spot was approximately $3$ to $4$. The ground state of the NV-center is a spin triplet ($^3A_2$), with its sublevels experiencing energy splitting due to spin-spin interactions, resulting in spin states with spin projection $|m_s = 0\rangle$ and $|m_s = \pm 1\rangle$, separated by $D = 2.87$ GHz when no external magnetic field is applied \cite{rondin2014magnetometry,doherty2013nitrogen}. In this context, $m_s$ refers to the spin projection along the intrinsic quantization axis of the NV-center, which corresponds to the direction towards the nitrogen atom and vacancy. The defect center can be optically excited via a spin-conserving transition to the excited level $^3E$, which is also a spin triplet. In addition, the $^3E$ excited state has an orbital doublet configuration that is effectively averaged at room temperature, resulting in a zero-field splitting of $D = 1.42$ GHz, maintaining the same quantization axis and gyromagnetic ratio as in the ground state \cite{rondin2014magnetometry,doherty2013nitrogen,maze2011properties}. After being optically excited to the $^3E$ level, the NV-center can relax back either by emitting photons through a radiative transition, which produces broad red ($600 - 800$nm) photoluminescence, or via a secondary pathway involving nonradiative intersystem crossing (ISC) to singlet states \cite{suter2017single,rondin2014magnetometry}. Experimental observations indicated the existence of two singlet states located between the ground and excited states of the NV-center, which are significant in the spin dynamics of the NV-center. Specifically, while optical transitions predominantly conserve spin ($\Delta m_s = 0$), non-radiative ISCs to the $^1E$ singlet state exhibit strong spin selectivity because the ISC transition rate from the $|m_s = 0\rangle$ sublevel is much lower than that from $|m_s = \pm 1\rangle$ \cite{suter2017single,rondin2014magnetometry}. In contrast, the NV-center favors the decay from the lowest $^1A_1$ singlet state to the ground state $|m_s = 0\rangle$. These spin-selective mechanisms result in significant electron spin polarization towards the $|m_s = 0\rangle$ state, achieved through optical pumping. Moreover, because ISCs are nonradiative processes, the intensity of the NV-center PL is considerably enhanced when the $|m_s = 0\rangle$ state is populated \cite{suter2017single,rondin2014magnetometry}. This spin-dependent response in PL allows the detection of electron spin resonance (ESR) or optically detected magnetic resonance (ODMR) at the level of individual defects through an optical readout \cite{suter2017single,rondin2014magnetometry}. When a single NV-center that was initially prepared in the $|m_s = 0\rangle$ state via optical pumping was subjected to a resonant microwave field, resulting in a transition to the $|m_s = \pm 1\rangle$ spin state, a decrease in the PL signal was detected. The simplest ground-state Hamiltonian of the NV-center is,
\begin{equation*}
    H = hDS^2_z + hE(S^2_x - S^2_y) + g\mu_B\vec{B} \cdot \vec{S} 
\end{equation*}
\begin{equation}
    \Rightarrow H = hDS^2_z + g\mu_BB_{NV}S_z + g\mu_B(B_xS_x + B_yS_y)+hE(S^2_x - S^2_y) 
\end{equation}
where $h$ is the Planck's constant, $D$ is the zero-field splitting factor, $g$ is the  Landé g-factor of the NV-center electronic spin, $\mu_B$ is the Bohr magneton, and $B_{NV}$ is the magnetic field along the NV center quantization axis \cite{suter2017single,rondin2014magnetometry}. The off-axis zero-field splitting parameter $E$ results from local strain in the host diamond matrix. In the weak magnetic field amplitudes, the eigen-energy of the ground state Hamiltonian or the ESR frequencies of the NV-center,
\begin{equation}   
\nu_\pm (B_{NV}) = D \pm \sqrt{(\frac{g\mu_B}{h}B_{NV})^2 + E^2}. \end{equation}
At a zero bias field ($B_{NV} = 0$), the ESR frequencies are $\nu_\pm = D\pm E$ \cite{suter2017single,rondin2014magnetometry}. In most cases, symmetric splitting of the zero-field ODMR spectrum in a bulk crystal sample has been reported \cite{rondin2014magnetometry,udvarhelyi2018spin}.\\
Here, asymmetric splitting of the zero-field ODMR peak for both the nitrogen-ion-implanted and diamond nanopillar samples was observed. It is well known that ion implantation damages the lattice structure of diamond crystals, and a high-temperature annealing process is generally \textcolor{black}{implemented} to generate NV-centers, which also repair lattice damage to a certain extent \cite{favaro2016toward,lobaev2017influence,balmer2009chemical}. Similarly, in nanofabrication processes, the ICP-RIE process is \textcolor{black}{utilized} to etch the diamond crystal, which also damages the diamond lattice. Owing to the deformation of the lattice structure, asymmetric splitting in the zero-field ODMR spectrum was observed, as shown in Figures \ref{fig4}(d) and \ref{fig5}(b) \cite{alam2024determining}. \textcolor{black}{In Section~\ref{subsection-2.3}, the mathematical model developed by M. Sahnawaz Alam et al. \cite{alam2024determining} is briefly described to explain the asymmetric nature of the zero-field CW-ODMR spectrum arising due to the presence of in-plane shear strain in the diamond lattice.}

%%%%%%%%%%%%%%%%%%%%%%%%%%%%%%%%%%%%%%%%%%%%%%%%%%%%%%%%%%%%%%%%%%%%%%%%

\subsection{Discussion}
\label{subsection-2.3}

\textcolor{black}{In Subsection~\ref{subsection-2.2}, the zero-field CW-ODMR data for the two different diamond samples are presented. During ion implantation, energetic ions penetrate the diamond lattice and displace carbon atoms from their lattice sites, leading to the formation of vacancies, interstitials, and other lattice defects. These defects locally distort the sp$^3$-bonded structure, and the size mismatch between the implanted atoms and carbon further induces strain within the crystal lattice \cite{liang2025strain}. For the diamond nanopillar fabrication, diamond etching was carried out using the inductively coupled plasma reactive ion etching (ICP-RIE) technique. In this process, high-energy ions accelerated by the radio-frequency (RF) bias remove carbon atoms from the diamond lattice. Carbon–carbon bonds on the diamond surface are dissociated by irradiated oxygen, and the carbon atoms are removed through the formation and desorption of gaseous carbon monoxide and carbon dioxide molecules \cite{xu2021different}. Continuous ion bombardment introduces defects and dislocations, thereby generating additional strain in the crystal lattice.}\\
\textcolor{black}{The type of strain present in the diamond lattice can be identified by examining the strain tensor parameters. The strain model described in Ref.~\cite{alam2024determining} is discussed here to understand the nature of strain in both the ion-implanted diamond sample and the diamond nanopillar sample. The Hamiltonian for the ground state of an NV-center in the absence of a static magnetic field can be written as
\begin{equation}
    H = hDS^2_z + H_{strain}
\end{equation}
\begin{equation}
    \frac{H_{strain}}{h} = M_zS^2_z + M_x(S^2_y - S^2_x) + M_y(S_xS_y+S_yS_x) + N_x(S_xS_z + S_zS_x) + N_y(S_yS_z + S_zS_y)
\end{equation}
where $h$ is Planck's constant, $D=2.87$ GHz is the zero-field splitting parameter, $S_i$ ($i\in \{x,y,z\}$) are the spin-1 matrices, and $M_i$, $N_i$ ($i\in \{x,y,z\}$) denote the strain amplitudes \cite{barry2020sensitivity,alam2024determining,udvarhelyi2018spin}, The strain amplitudes are functions of $\epsilon_{ij}$ and $h_{ij}$, which are the strain tensor elements in the single NV-center frame, and the spin–strain coupling parameters \cite{barry2020sensitivity,alam2024determining,udvarhelyi2018spin}, respectively. The eigen-energies of the ground-state Hamiltonian in the absence of the microwave driving field are given by \cite{barry2020sensitivity,alam2024determining,udvarhelyi2018spin}
\begin{equation}
    E_{|m_{s} = 0\rangle} = 0, \qquad 
    E_\pm = \left(D + M_z \pm \sqrt{M^2_x + M^2_y}\right).
\end{equation}
The strain at the NV-center location mixes the $|\pm1\rangle$ spin states and generates new eigenstates of the NV-center Hamiltonian. In the absence of a static magnetic field, two microwave-active transitions appear in the ODMR spectrum with transition frequencies
\begin{equation}
\nu_\pm = \frac{1}{h}\left(D+M_z \pm \sqrt{M^2_x + M^2_y}\right)
\end{equation}
\cite{alam2024determining,udvarhelyi2018spin}. Each transition results in a depletion of the ground-state population, thereby reducing the fluorescence and producing two dips in the ODMR spectrum. Consequently, a double-peak Lorentzian function is suitable for fitting the zero-field ODMR spectrum, and the extracted parameters are directly related to the strain components.}\\
\textcolor{black}{For shear strain in XY-plane (where NV-center axis is the Z-axis), the strain tensor elements $\epsilon_{xx}$, $\epsilon_{yy}$, and $\epsilon_{xy}$ are non-zero, while $\epsilon_{zz}=0$. The component $\epsilon_{xx}$ contributes to the shift of the center frequency; however, no such shift is observed in our measurements. Therefore, $\epsilon_{yy}$ and $\epsilon_{xy}$ are considered non-zero, while $\epsilon_{xx}=\epsilon_{zz}=0$. The asymmetry in the ODMR spectrum arises from the non-zero $\epsilon_{xy}$ shear strain component present in the samples.}\\
\textcolor{black}{Based on the fitting parameters (see Fig.~\ref{figS2}) extracted from Figs.~\ref{fig4}(d) and~\ref{fig5}(b), the transverse strain was estimated to be $1.9875$~MHz for the nitrogen ion–implanted sample and $1.8835$~MHz for the diamond nanopillar sample. The spectral imbalances are 0.058 and 0.068 for the nitrogen ion–implanted sample and the diamond nanopillar sample, respectively.}
%%%%%%%%%%%%%%%%%%%%%%%%%%%%%%%%%%%%%%%%%%%%%%%%%%%%%%%%%%%%%%%%%%%%%%%%

\section{Conclusion}
\label{section-3}

\textcolor{black}{In conclusion, asymmetric splitting in the zero-field CW-ODMR spectrum of NV-centers was observed in both nitrogen–ion–implanted diamond and diamond nanopillar samples, arising from shear strain in the crystal lattice. Identifying and quantifying strain in diamond is an important characterization step for understanding qubit performance in applications involving atomic-scale quantum sensors and quantum network nodes.}\\
\textcolor{black}{For quantum communication, particularly for transferring quantum information to spin qubits or generating entanglement between distant qubits via single-photon detection schemes, efficient light-matter interaction plays a crucial role \cite{ruf2021quantum}. Similarly, high photon collection efficiency plays an important role in quantum information processing and quantum sensing, as it directly impacts the overall system performance \cite{ruf2021quantum}. Ion implantation and nanofabrication of diamond photonic structures therefore serve as key enabling techniques for improving these capabilities. Ion implantation followed by annealing leads to the formation of NV-centers within the diamond lattice, typically with minimal lattice disorder at low implantation doses. However, higher ion doses can introduce significant lattice distortion, resulting in splitting and asymmetry in the ODMR spectrum.}\\
\textcolor{black}{Zero-field ODMR of NV-centers provides a powerful method for probing strain components in diamond crystals, offering insights into qubit coherence \cite{van2019optical}, device performance, and the calibration of quantum gates \cite{PhysRevApplied.23.034052}, as well as the optical properties of the emitted photons \cite{maze2011properties}. Consequently, this approach provides a valuable framework for evaluating and optimizing device performance in quantum sensing, quantum information processing, and quantum communication applications.}
%%%%%%%%%%%%%%%%%%%%%%%%%%%%%%%%%%%%%%%%%%%%%%%%%%%%%%%%%%%%%%%%%%%%%%%%

\section{Experimental Details}
\label{section-4}

\textcolor{black}{This section describes the measurement setup in detail. The Raman spectroscopy setup is introduced first, followed by the optical and electronic components of the home-built microwave-excitation-integrated laser-scanning confocal system.}

\subsection{Raman Spectroscopy}
\label{section-4.1}

The ion-implanted samples were characterized using a confocal Micro-Raman spectrometer (Horiba Jobin Yvon LAB RAM HR Evolution) in backscattering geometry (resolution $~0.1$ cm$^{-1}$), employing a $532$ nm diode laser as the excitation source. For quantitative impurity analysis, optical absorbance spectra were recorded using a UV–Visible spectrophotometer (Carry $100$–UV–Vis from Agilent).

\subsection{Microwave Integrated Laser-Scanning Confocal Microscope}
\label{section-4.2}

\begin{figure}[h!]
\centering
\includegraphics[width=12cm]{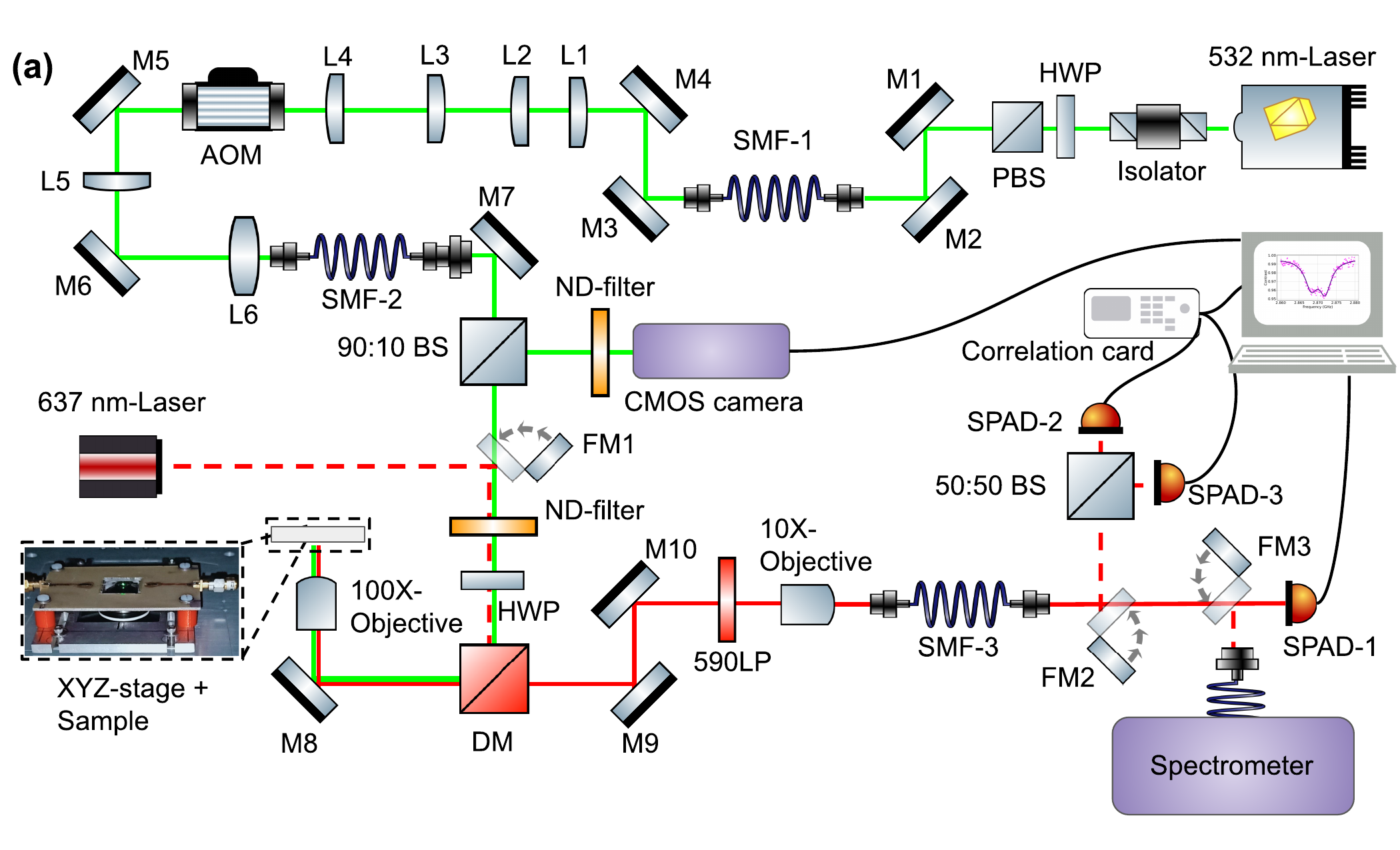}
\includegraphics[width=12cm]{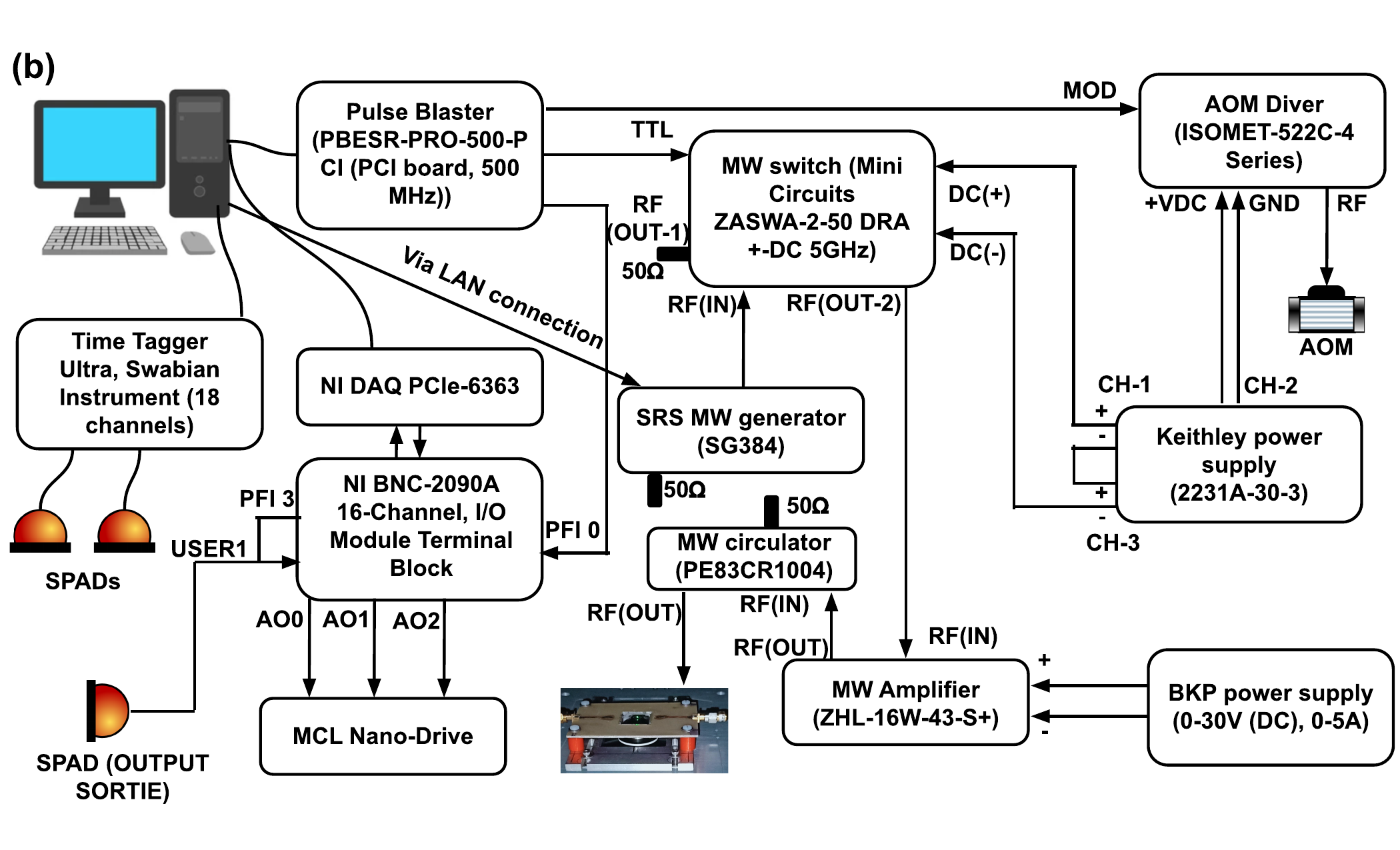}
\caption{(a)Schematic diagram of the home-built laser scanning confocal microscope with microwave excitation setup. Abbreviation-L: Plano convex lens, M: Broadband dichroic mirrors, HWP: Half-wave plate, PBS: Polarizing beam splitter, SMF: Single-mode fiber, AOM: Acousto-optic modulator, ND-filter: Neutral density filters, DM: Dichroic mirror, LP: Long-pass filter, FM: Flip mirror, BS: Beam splitter, SPAD: Single-photon avalanche diode. (b) Electronics for controlling the laser-scanning confocal imaging setup and spin-qubit-based experiments.}
\label{fig6and7}
\end{figure}

A laser-scanning confocal microscope with microwave excitation was used for strain measurement in the diamond samples. A continuous-wave (CW) $532$ nm diode-pumped solid-state (DPSS) laser (Sprout-G-5W from Lighthouse Photonics) was used for the optical pumping. A half-wave plate (HWP) and polarizing beam splitter (PBS) combination controls the laser power. An acoustic-optic modulator (M1133-aQ80L-1 from Isomet) was added to the excitation path to pulse the $532$ nm laser. A single-mode fiber (P3-405B-FC-1 from Thorlabs) was used for spatial filtering to create a clean Gaussian $532$ nm excitation beam. The collimated, clean Gaussian was sent to the 90:10 beam splitter (BS), followed by a dichroic mirror ($532$ nm laser BrightLine dichroic beam splitter from Semrock) to separate the excitation beam and the NV-center emission spectra. A $590$ nm long-pass filter (ET590lp from Chroma) was used for spectral filtering and to achieve a better signal-to-noise ratio. A 10X objective lens is used to focus down the beam in the detection path, and a single-mode fiber is placed at the focal spot of the objective lens, which acts as a pinhole system for detecting light from the focal spot of the 100X objective lens inside the diamond crystal sample. In the detection path, a single-photon avalanche diode (SPCM-AQRH-4X from Excelitas) was used for imaging and PL collection for ODMR measurement. A flip mirror couples the Hanbury Brown and Twiss (HBT) setup for second-order correlation measurements (correlation card Time Tagger Ultra from Swabian Instruments) or anti-bunching measurements. A three-axis piezo nanopositioning stage (Nano-LP200, Mad City Labs Inc.) was used to scan the diamond crystal and diamond nanopillar sample. Figure \ref{fig6and7}(a) shows a schematic of the laser scanning confocal microscope with the microwave excitation setup.
\textcolor{black}{Figure~\ref{fig6and7}(b) illustrates the schematic of the electronic control setup used for confocal imaging and spin-qubit experiments. A pulse blaster (PBESR-PRO-500-P) generates synchronized timing signals to the NI DAQ PCIe-6363 card and the acousto-optic modulator (AOM) driver, while also controlling the microwave (MW) switch (Mini-Circuits ZASWA-2-50 DRA). The NI DAQ, interfaced with a BNC-2090A 16-channel terminal block, provides analog outputs (AO0–AO2) to drive the Mad City Labs (MCL) Nano-Drive for precise positioning and raster scanning, and acquires input signals from the single-photon avalanche detector (SPCM-AQRH-16-FC) through the PFI3 DAQ channel. Microwave (MW) signals are produced by an SRS SG384 source and passed through a circulator before being amplified by a ZHL-16W-43-S$+$ power amplifier, with bias supplied by a BK Precision power supply. The amplified MW signals are then delivered to the sample via an antenna. The acousto-optic modulator (AOM, ISOMET-522C-x series) is driven by an AOM driver powered by a Keithley (2231A-30-3) supply, enabling optical modulation of the excitation laser. Additionally, a CMOS camera is used in parallel to focus the sample by imaging the diffraction spot of the excitation beam.}

%%%%%%%%%%%%%%%%%%%%%%%%%%%%%%%%%%%%%%%%%%%%%%%%%%%%%%%%%%%%%%%%%%%%%%%%

\section*{Acknowledgements}

K. Saha acknowledges financial support from the National Quantum Mission (NQM) and the Chanakya Doctoral Fellowship for A. Majumder and A. Bathla. Authors also acknowledge the IIT Bombay Nanofabrication Facility (Centre of Excellence in Nanoelectronics) and the Sophisticated Analytical Instrument Facility (SAIF)/ Center for Research in Nano Technology and Science (CRNTS), IIT Bombay, for support in diamond nanopillar fabrication and optical characterization, respectively. Authors also acknowledge the insightful contributions of Dr. Bikash D. Choudhury for the initial fabrication of diamond nanopillars. P. Rai acknowledges the funding support from the SERB (CRG/2021/000696) and DAE (DPR-6/3(1)/2018/UM-DAE-CBS/R$\&$D-II/8971). The authors acknowledge the Solid State Physics Laboratory (SSPL), New Delhi, India, for the ion implantation facility. The authors would like to thank Alexander Franzen for creating the illustration of the optical components.
%%%%%%%%%%%%%%%%%%%%%%%%%%%%%%%%%%%%%%%%%%%%%%%%%%%%%%%%%%%%%%%%%%%%%%%%

\section*{Author contributions}

AM performed the setup development and ODMR experiments. VKS conducted the SRIM simulations and Raman spectroscopy measurements. AB fabricated the diamond nanopillars, while BSY and NS carried out the ion implantation. AM and VKS prepared the manuscript in consultation with KS and PR, who supervised all aspects of the work.
%%%%%%%%%%%%%%%%%%%%%%%%%%%%%%%%%%%%%%%%%%%%%%%%%%%%%%%%%%%%%%%%%%%%%%%%

\section*{Conflict of interest}

The authors declare no conflict of interest.
%%%%%%%%%%%%%%%%%%%%%%%%%%%%%%%%%%%%%%%%%%%%%%%%%%%%%%%%%%%%%%%%%%%%%%%%

\section*{Supporting Information}
\label{Supporting}

\subsection*{Diamond Nanopillar Simulation Details}
\begin{figure}[h]
\centering
\includegraphics[width=12cm]{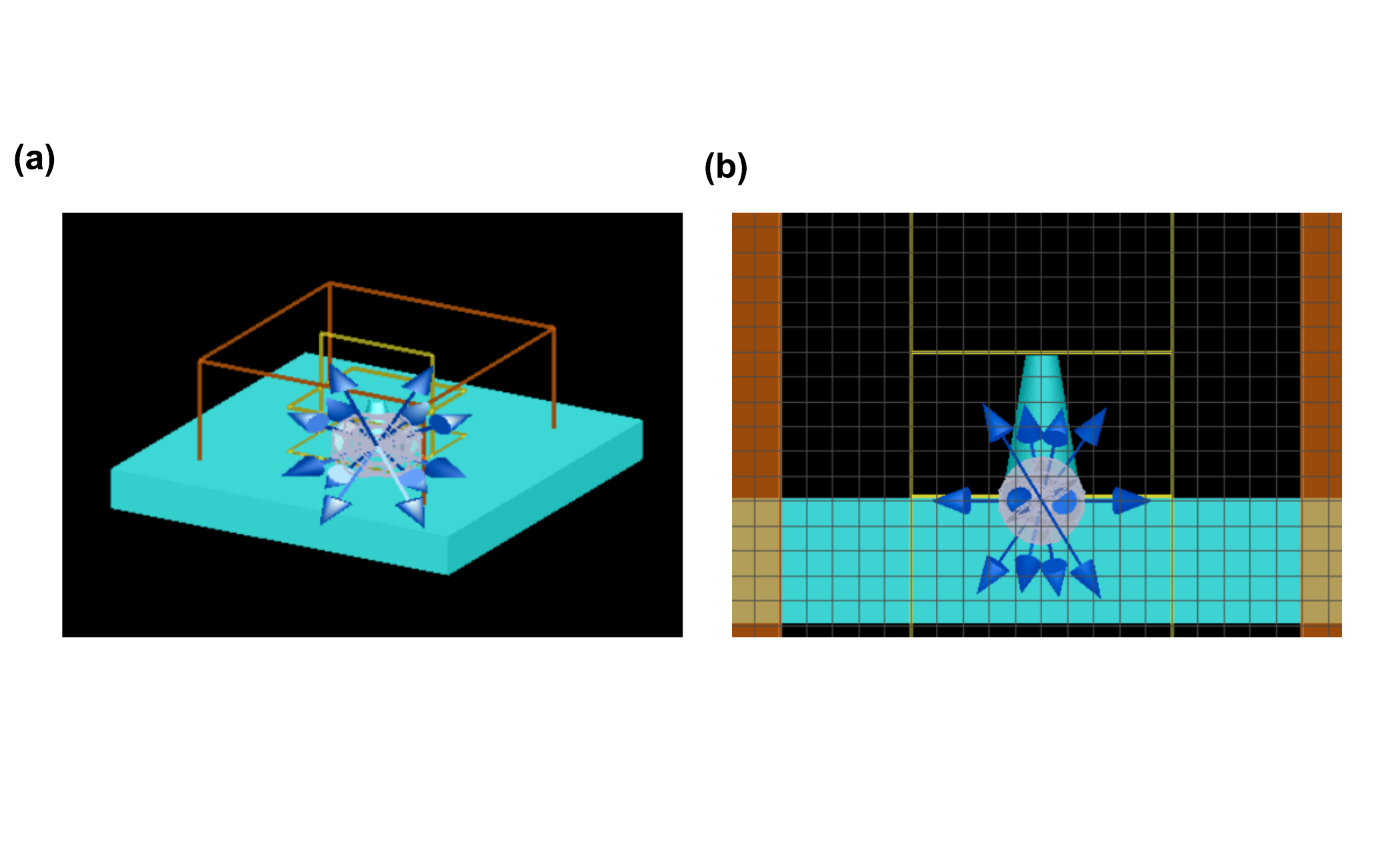}
\vspace*{-50pt}
\caption{\textcolor{black}{(a) 3D-FDTD simulation setup of a diamond nanopillar in Ansys Lumerical software. (b) XZ cross-sectional view of a diamond nanopillar with an ensemble of NV centers located at the base of the pillar.}}
\label{figS1}
\end{figure}

\textcolor{black}{A three-dimensional finite-difference time-domain (3D-FDTD) module was used to simulate the propagation and confinement of light in the diamond nanostructure. The simulations were performed at room temperature ($300$~K) using eight dipoles corresponding to the four possible orientations of the NV center, with a wavelength span of $1$~nm. A non-uniform mesh with a mesh accuracy of $2$ was employed. The time stability factor ($dt$) and the corresponding time step were set to $0.99$ and $0.493$~fs, respectively, with a minimum mesh step size of $2.5\times10^{-10}~\mu$m. Figure~\ref{figS1} illustrates the 3D-FDTD simulation setup of a diamond nanopillar implemented in the Ansys Lumerical software, along with the XZ cross-sectional view of the nanopillar containing an ensemble of NV centers located at the base of the structure. A frequency-domain power monitor was used to analyze the light confinement induced by the nanopillar.}
%%%%%%%%%%%%%%%%%%%%%%%%%%%%%%%%%%%%%%%%%%%%%%%%%%%%%%%%%%%%%%%%%%%%%%%%

\subsection*{NV-Center Electronic-Level Structure and ODMR Measurement Details}
\begin{figure}[h]
\centering
\includegraphics[width=12cm]{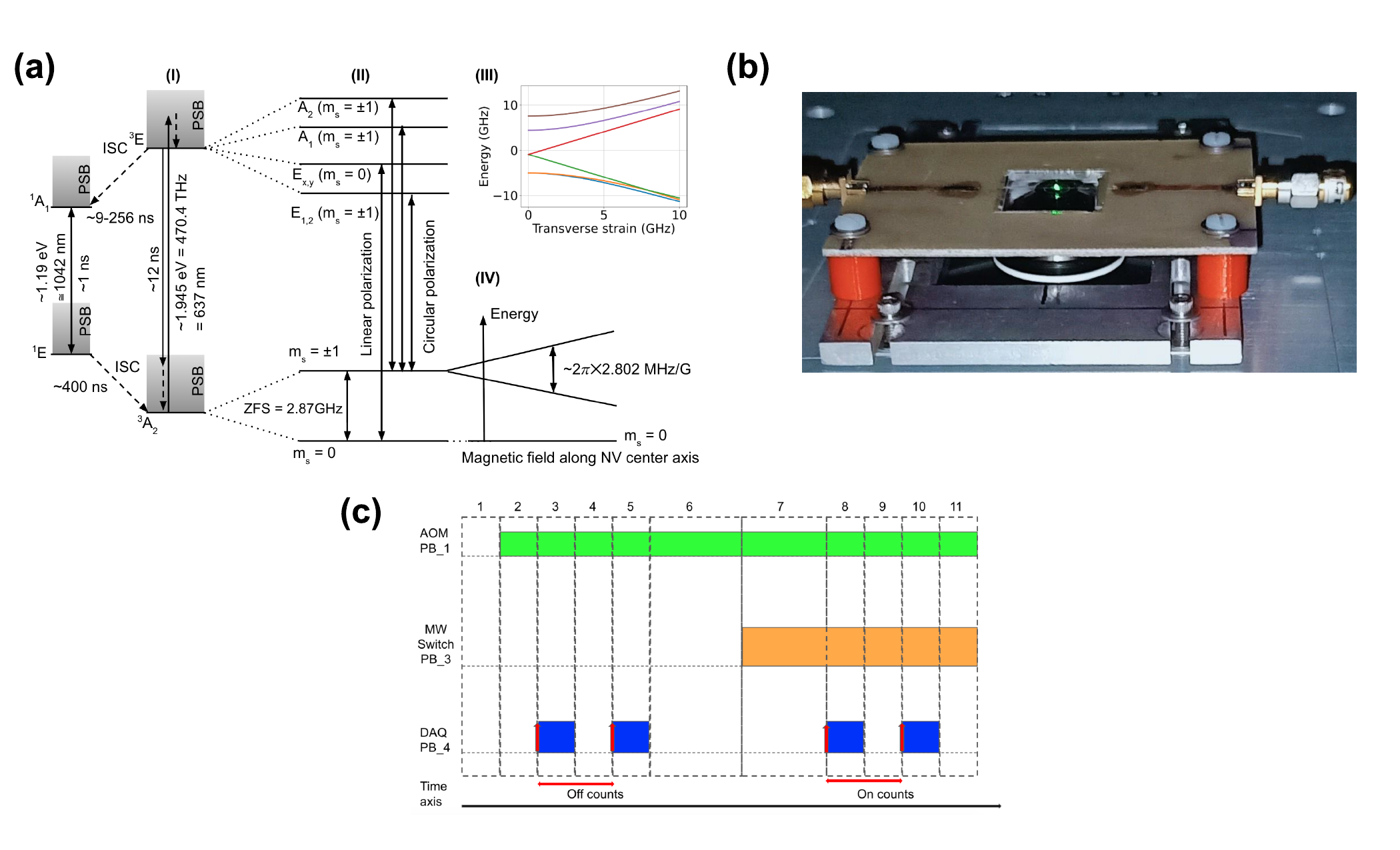}
\caption{\textcolor{black}{(a) Electronic energy-level structure of the NV-center in diamond. (b) Photograph of the microwave excitation module. (c) Control pulse sequence used for the CW-ODMR measurement.}}
\label{figS2}
\end{figure}

\begin{figure}[h]
\centering
\includegraphics[width=12cm]{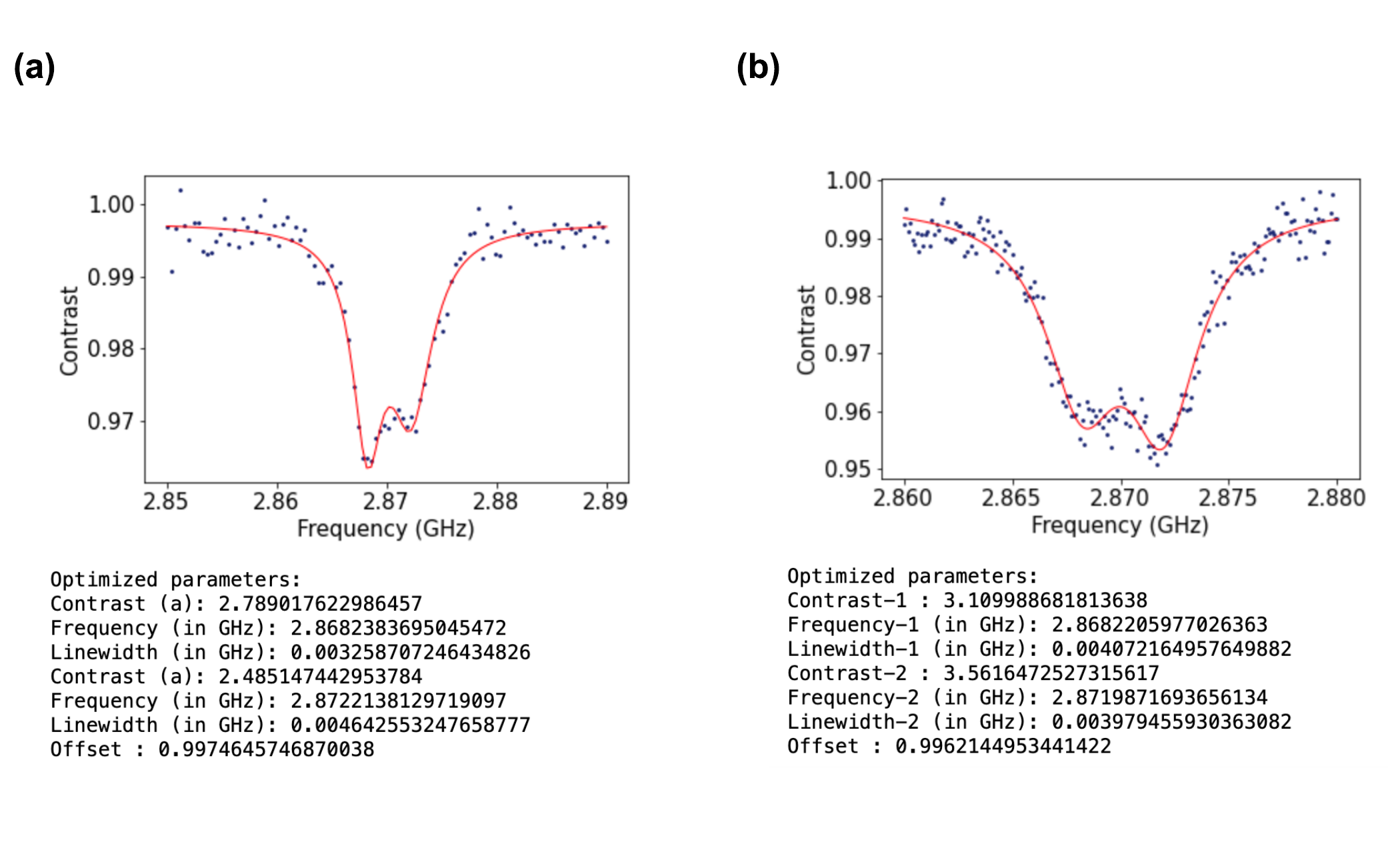}
\caption{\textcolor{black}{Fitted CW-ODMR spectra for (a) the nitrogen-ion-implanted diamond sample and (d) the diamond nanopillar sample. The experimental data are fitted using a double-peak Lorentzian curve.}}
\label{figS3}
\end{figure}

\textcolor{black}{Figure~\ref{figS2}(a) shows the electronic energy level diagram of the negatively charged nitrogen vacancy defect center in a diamond crystal. (I) The optical transitions between the ground-state triplet $^3A_2$ and the excited-state triplet $^3E$ are shown, along with the intermediate singlet states $^1A_1$ and $^1E$, which enable non-radiative intersystem crossing. The zero-phonon line (ZPL) at 1.945~eV (637~nm) and the phonon sidebands (PSB) are indicated, together with characteristic radiative and non-radiative lifetimes. (II) Fine structure of the excited state $^3E$, including spin–orbit and spin–spin interactions that split the $m_s = 0$ and $m_s = \pm1$ sublevels, with linear and circular polarization selection rules. (III) Effect of transverse strain or electric field on the $^3E$ manifold, leading to energy-level mixing and splitting of the $A_1$, $A_2$, $E_x$, and $E_y$ sublevels. (IV) Ground-state spin sublevels ($^3A_2$) showing the zero-field splitting ($D \approx 2.88$~GHz) between $m_s = 0$ and $m_s = \pm1$, and Zeeman splitting under an external magnetic field $B_z$ \cite{ruf2021cavity}. Figure~\ref{figS2}(b) shows the sample holder integrated with the microwave excitation module. A thin copper wire with a diameter of $25~\mu m$ was used to deliver the microwave field required for driving the spin transitions. Figure~\ref{figS2}(c) presents the pulse sequence generated by the pulse blaster for the CW-ODMR experiment. The PB-1, PB-3, and PB-4 channels of the pulse bluster module were used to control the AOM (for pulsing the laser), MW switch (for pulsing the MW excitation field), and NI-DAQ (for photon counting), respectively. An edge-counting protocol was employed to measure the photon counts. The readout time was set to $300$~ns, corresponding to the separation between two rising edges. The off-counts $(C_{off})$ and on-counts $(C_{on})$ represent the photon counts measured without and with the microwave excitation pulse, respectively. The contrast was calculated using the formula $C = (C_{off}-C_{on})/(C_{off}+C_{on})$. The microwave signal, generated by the SRS-SG source, had a power of $-12$~dBm. The laser power was 1.3~mW.}
\textcolor{black}{Figure~\ref{figS3} shows the zero-field CW-ODMR data along with the fitted parameters for the nitrogen ion-implanted CVD-grown diamond sample and the nanopillars fabricated on a CVD-grown diamond sample. A double-peak Lorentzian function was used to fit the experimental data. The resonance frequencies, linewidths, and contrasts were extracted from the fitted parameters, as shown in Figures~\ref{figS3}(a) and (b).}
%%%%%%%%%%%%%%%%%%%%%%%%%%%%%%%%%%%%%%%%%%%%%%%%%%%%%%%%%%%%%%%%%%%%%%%%

\printendnotes

% Submissions are not required to reflect the precise reference formatting of the journal (use of italics, bold etc.), however it is important that all key elements of each reference are included.
%\bibliography{main}

\end{document}